\documentstyle[prb,aps,multicol]{revtex}
\input{psfig}

\def\be{\begin{equation}}
\def\ee{\end{equation}}
\def\ba{\begin{eqnarray}}
\def\ea{\end{eqnarray}}

\begin{document}

\title{Finite voltage shot noise in normal-metal -- superconductor junctions}

\author{Alban~L.\ Fauch\`ere$^{a}$, Gordey~B.\ Lesovik$^{b}$, 
and Gianni Blatter$^{a}$}
\address{$^{a}$Theoretische Physik, Eidgen\"ossische Technische Hochschule,
 CH-8093 Z\"urich, Switzerland\\
$^{b}$Institute of Solid State Physics, Chernogolovka 142432,
 Moscow district, Russia }

\date{March 13, 1998}

\maketitle

\begin{abstract}
We express the low-frequency shot noise in a disordered normal-metal -- 
superconductor (NS) junction at finite (subgap) voltage in terms of the 
normal scattering amplitudes and the Andreev reflection amplitude. 
In the multichannel limit, the conductance exhibits resonances which are 
accompanied by an enhancement of the (differential) shot noise. In the study of 
multichannel single and double barrier junctions we discuss the noise 
properties of coherent transport at low versus high voltage with respect to 
the Andreev level spacing.
\end{abstract}

\begin{multicols}{2}

Shot noise arises from the current fluctuations in transport as a 
consequence of the discrete nature of the carriers, first predicted
by Schottky for the vaccuum tube\cite{schottky}. In a coherent 
conductor the quantum fluctuations, which follow from the probabilistic
nature of the backscattering restricted by the Pauli principle, are 
reduced in comparison to the Schottky result\cite{lesovik}.
In a disordered normal-metal -- superconductor (NS) junction, the shot noise 
is produced by the normal scattering processes as well as the imperfect 
Andreev reflection\cite{khlus,muzy,Jong,hessling}. The Andreev reflection 
introduces fluctuations proportional to the double electron charge, which may 
be iteratively increased to give fluctuations of several charge quanta in 
biased SNS junctions\cite{averimam}.
In this paper we study the shot noise at finite voltage in a disordered
NS junction as well as its  structure due to the iterative 
scattering processes between the disordered normal lead and the NS interface.
We derive a general formula which expresses the differential shot noise of the
dirty NS junction in terms of the scattering matrix of the normal
lead and the (scalar) amplitude of Andreev reflection. In the models of a 
single barrier normal-metal -- insulator -- normal-metal -- superconductor 
(NINS) junction and a double barrier NINIS junction, we explain the existence 
of a non-trivial resonance structure.

\begin{figure}[tb]
\vspace{-3mm}
\centerline{\psfig{figure=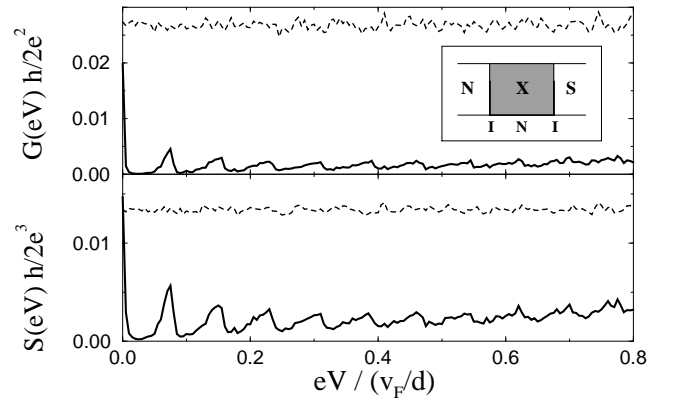,angle=-90,width=80mm}}
\narrowtext
\vspace{-1.2cm}
\caption{Differential conductance (top) and shot noise (bottom) in a 
multichannel NINIS junction (see inset, symmetrical barriers of strength 
$\int dx\, V\left(x\right) = 
3\hbar v_F$, $T_1=T_2 \approx 0.05$, interbarrier distance 
$d=20\, v_F/\Delta$, $8\times 10^4$ channels). The average shot 
noise per channel [Eq.\ (\ref{diffnoise})] exhibits maxima at 
the resonances of the conductance (solid lines). Note the enhanced structure 
of the noise with respect to the conductance [Eq.\ (\ref{cond})]. 
The dashed lines show the corresponding results for the normal state NININ 
junction.
Inset: schematic NXS junction, e.g., X $=$ INI. }
\label{fig1}
\end{figure}

We consider a disordered NXS junction, shown in the inset of Fig.\ \ref{fig1}, 
with an arbitrary elastic scattering 
region X in the normal lead, whose effect on the noise is to be determined.
The low-frequency power spectrum of the current fluctuations is determined by 
the irreducible current--current correlator 
\be P =\int dt\, e^{i\omega t} \langle\langle 
I\left( t\right) I\left( 0\right) \rangle\rangle, \quad \quad   
\omega\to 0 
\label{power} \ee
($\langle\langle I\left(t\right) I\left(0\right) \rangle
\rangle  = \langle  \left(I\left(t\right)-\langle I\rangle\right) 
\linebreak[0] 
\left(I\left( 0\right)-\langle I\rangle \right) \rangle$ is the second 
cumulant in time). The time-dependence of the current operator is defined 
by 
\be I\left(t\right)=
\exp \left[i\left(H-\mu N\right)t\right]\, I \,\exp \left[-i\left(H-
\mu N\right)t\right].
\ee
In the mean field approximation for the Hamiltonian $H$
we do not account for the fluctuations of the order parameter in the S region. 
The effective Hamiltonian is diagonalized by a Bogoliubov 
transformation\cite{kuemmel}.
After integration over the cross-section, the net current operator can be 
expressed through the positive energy eigenfunctions,
\ba I\left(t\right) = \frac{-e}{mi}\sum_{\epsilon_m,\epsilon_n>0} 
\int\! dydz  && \big(u_m^*\hat{\partial_x} 
u_n \gamma_{m}^{\dag}\gamma_{n} \nonumber \\ && 
- v_m^* \hat{\partial_x} v_n 
\gamma_{n}\gamma_{m}^{\dag} \big) 
e^{i\left(\epsilon_m-\epsilon_n\right)t} 
\label{currentop}
\ea
which may be evaluated in the normal lead. The operators $\gamma_{m}$ 
belonging to the wavefunctions $\left(u_m,v_m\right)$ annihilate the 
scattering states of the disordered NS junction 
(note $u\hat{\partial_x}v=u\partial_xv-v\partial_xu$). We note that
the symmetry of the BdG eigenfunctions $(u_n,v_n)$, $\epsilon_n$ with 
respect to sign reversal of energy to $(-v_n^*,u_n^*)$, $-\epsilon_n$ 
is the consequence of spin degeneracy in Nambu space, allowing the 
the negative energy states to be eliminated in favor of the positive
energy states.
\noindent 
\noindent
At voltages below the superconducting gap ($eV<\Delta$), 
the quasi-particles are injected from the normal reservoir only, and the 
wavefunctions depend on the $N\times N$ reflection matrices 
$r_{ee}\left(\epsilon\right)$, $r_{he}\left(\epsilon\right)$, 
$r_{eh}\left(\epsilon\right)$, and $r_{hh}\left(\epsilon\right)$ of the 
entire disordered NS junction \cite{notation}. The scattering states indexed by 
$m=\left[^{e}_{h}, \mu,\epsilon\right]$ consist of an incident $e$(lectron) or $h$(ole) 
like quasi-particle in channel $\mu$ with energy $\epsilon$ superposed with the 
reflected electron and hole states. The occupation numbers are given by 
the Fermi-Dirac distribution $f_{e,h}=f\left(\epsilon \pm eV\right)$ 
(the voltage is measured with respect to the chemical potential of the 
superconductor).

The noise power (\ref{power}) is determined by the transitions 
induced by the current operator $I$ from initial states 
$|n\rangle = |n_n=1, n_m=0\rangle$ to intermediate states 
$|m\rangle = |n_n=0, n_m=1\rangle$ and back, which differ only by their 
occupation with respect to two single particle states with indices $n$ and $m$.
E.g., the transition between an incident electron 
$n = \left[e,\nu,\epsilon\right]$ and an incident hole 
state $m=\left[h,\mu,\epsilon\right]$ is produced by the
matrix element between the reflected electrons and holes of the two scattering 
states with respect to the current operator, 
$\langle m | I | n\rangle \propto f_{e} \left(1-f_{h}\right) 
 (r_{eh}^{\dag}r_{ee}-r_{hh}^{\dag}r_{he})_{\mu\nu}$.
Summed over the channels, the transitions contribute to the fluctuations 
with the weight 
$\sum_{\mu,\nu} |\langle m | I | n\rangle|^2 \propto  
f_{e}\left(1-f_{h}\right) \mbox{Tr}\{
(r_{ee}^{\dag}r_{eh}-r_{he}^{\dag}r_{hh}) \linebreak[0]
(r_{eh}^{\dag}r_{ee}-r_{hh}^{\dag} r_{he}) \} \linebreak[1] 
= f_{e}\left(1-f_{h}\right) \mbox{Tr}\{r_{he}^{\dag}r_{he}
(1-r_{he}^{\dag}r_{he})\}$.
Following the quantum mechanical formalism outlined, we obtain the 
low-frequency limit of the power spectrum\cite{datta} valid for 
$T\ll eV < \Delta$,
\end{multicols}
\widetext
\be \!\!\!\!\! P = \frac{4e^2}{h} \int\limits_{0}^{\Delta} d\epsilon \left\{ 
\left[ f_{e}\left(1-f_{h}\right) + f_{h}\left(1-f_{e}\right) 
\right] 
\mbox{Tr}\Big[ r_{he}^{\dag}r_{he}\left(1-r_{he}^{\dag}r_{he}\right)
\Big]
+\left[ f_{e}\left(1-f_{e}\right) + f_{h}\left(1-f_{h}\right) \right] 
\mbox{Tr}\Big[ \left(r_{he}^{\dag}r_{he}\right)^2\Big] \right\}. 
\label{pgeneral} \ee
\vspace{-3mm}
\begin{multicols}{2}
\noindent
The first term describes the transitions between states of opposite current 
sign, while the second represents the transitions 
between states of equal current sign. At zero temperature, the second term 
is suppressed by the Pauli principle and the first term produces the 
shot noise\cite{lesovik,lesolevi}. 
The thermal fluctuations describing the Johnson-Nyquist noise\cite{Nyquist} 
are due to both terms. Eq. (\ref{pgeneral}) is manifestly invariant under 
sign reversal of voltage\cite{leso}.

{\it Interplay between normal scattering and Andreev reflection.}
We express the electron-hole reflection matrix $r_{he}\left(\epsilon\right)$ 
in Eq.\ (\ref{pgeneral}) in terms of the normal reflection- and transmission 
matrices $r_{ii}$ and $t_{ij}$ of the scattering (X) region and the amplitude 
of Andreev reflection $\Gamma$ at the XS interface,
$r_{he}\left(\epsilon\right)=  t_{12}^*\left(-\epsilon\right) \linebreak[0] 
\left[1-\Gamma^2\left(\epsilon\right) r_{22}\left(\epsilon\right) 
r_{22}^*\left(-\epsilon\right) \right]^{-1} \linebreak[0] 
\Gamma\left(\epsilon\right)  t_{21}\left(-\epsilon\right)$
(scattering matrices for $X$, see Fig.\  \ref{fig1}: 
$r_{ii}\left(\epsilon\right)$, $t_{ij}\left(\epsilon\right)$ for electrons, 
$r_{ij}^{*}\left(-\epsilon\right)$, $t_{ij}^{*}\left(-\epsilon\right)$ 
for holes, $i,j=1 (2)$ left (right), 
$\Gamma\left(\epsilon\right)=e^{i\vartheta\left(\epsilon\right)}$, 
$\vartheta\left(\epsilon\right)=\arccos\left(\epsilon/\Delta\right)$).
Inserting $r_{he}\left(\epsilon\right)$ in Eq.\ \ref{pgeneral} provides us
with a general multichannel expression for the shot noise in a disordered
NXS junction, explicit in the normal scattering matrix and the Andreev 
reflection amplitude.

Next we restrict ourselves to the situation of a junction with uniform
transverse cross-section, which permits the separation of channels and
will be used in the single and double barrier systems below.
We recall\cite{leso} that the differential conductances of the channels $\nu=1 ... N$ 
are given by the matrix product $r_{he}^{\dag}r_{he}$ in Eq.\ (\ref{pgeneral}),
\end{multicols}
\be \frac{h}{4e^2} G_{\nu}\left(\epsilon\right)= \left(r_{he}^{\dag}r_{he}
\right)_{\nu\nu} = \frac{ T_{\nu}\left(\epsilon\right) T_{\nu}\left(-\epsilon
\right)} 
{T_{\nu}\left(\epsilon\right) T_{\nu}\left(-\epsilon\right) + 
R_{\nu}\left(\epsilon\right) + R_{\nu}\left(-\epsilon\right) - 
2 {\rm Re}\left[\Gamma \left(\epsilon\right)^2 r_{\nu}\left(\epsilon\right) 
r_{\nu}^*\left(-\epsilon\right)\right]}. \label{cond}
\ee
\vspace{-3mm}
\begin{multicols}{2}
\noindent
$r_{\nu}\left(\epsilon\right)$ is the reflection amplitude of channel $\nu$ in 
the matrix $r_{22}\left(\epsilon\right)$ describing the reflection from the 
{\it right}-hand side of the X region, $R_{\nu}\left(
\epsilon\right)=|r_{\nu}\left(\epsilon\right)|^2$, and $T_{\nu}\left(\epsilon
\right)=1-R_{\nu}\left(\epsilon\right)$. 
The denominator of Eq. (\ref{cond}) contains the 
interference term describing the iterative scattering processes between the 
X region and the XS interface. At zero temperature, the shot noise (\ref{pgeneral}) 
exhibits the noise power $P = \frac{1}{e} \int_0^{e|V|} d\epsilon\, \sum_{\nu} 
S_{\nu}\left(\epsilon\right)$ with the differential (low-frequency) 
shot noise of channel~$\nu$,
\end{multicols}
\be \frac{h}{2e^3}S_{\nu}\left(\epsilon\right) = 4
\left(r_{he}^{\dag}r_{he} \left(1-r_{he}^{\dag}r_{he}\right)\right)_{\nu\nu} =
\frac{ 4 T_{\nu}\left(\epsilon\right) T_{\nu}\left(-\epsilon\right)
\left(R_{\nu}\left(\epsilon\right) + R_{\nu}\left(-\epsilon\right) -
2 {\rm Re}\left[\Gamma \left(\epsilon\right)^2 r_{\nu}\left(\epsilon\right) 
r_{\nu}^*\left(-\epsilon\right)\right]\right)} 
{\left(T_{\nu}\left(\epsilon\right) T_{\nu}\left(-\epsilon\right) + 
R_{\nu}\left(\epsilon\right) + R_{\nu}\left(-\epsilon\right) - 
2 {\rm Re}\left[\Gamma \left(\epsilon\right)^2 r_{\nu}\left(\epsilon\right) 
r_{\nu}^*\left(-\epsilon\right)\right]\right)^2}.
\label{diffnoise} \ee
\vspace{-3mm}
\begin{multicols}{2}
\noindent
The two energy dependencies of the reflection probabilities 
$R_{\nu}\left(\epsilon\right)$ and (the phases of) the reflection amplitudes
$r_{\nu}\left(\epsilon\right)$ translate into the voltage dependence 
of the shot noise. The energy dependence of the Andreev amplitude can often
be neglected. In the limit $\epsilon\to 0$ ($\Gamma\to -i$) the 
phase dependencies drop out of Eq.\ (\ref{diffnoise}) and we recover 
the linear response result\cite{deJong} 
$\left(h/2e^3\right)S_{\nu}\left( 0\right)= 
16 T(0)^2 [1-T(0)]/[2-T(0)]^4$. 

\begin{figure}[htb]
\vspace{-5mm}
\centerline{\psfig{figure=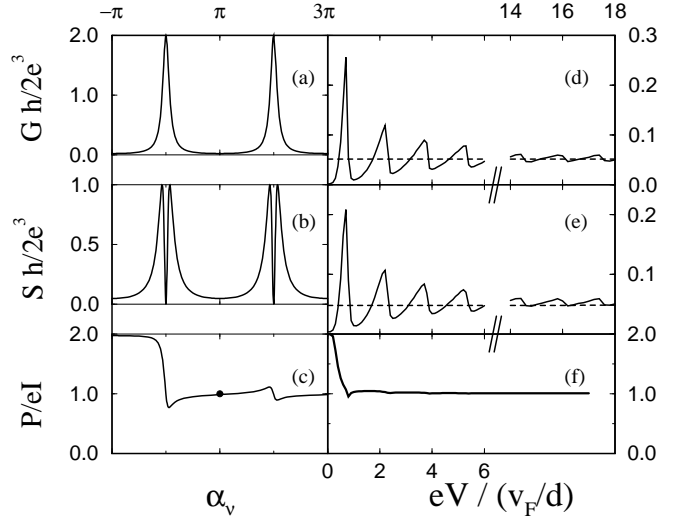,width=90mm,angle=-90}}
\narrowtext
\caption[*]{Zero temperature conductance (a,d), differential shot noise (b,e), 
and finite voltage noise power to current ratio (c,f) for a NINS junction with a 
barrier of strength $\int dx\, V(x) = 3\hbar v_F$, average transmission 
$T=0.05$, 
$\epsilon_F=500\Delta$, $d=20\, v_F/\Delta$ 
($v_F/d \ll \Delta \ll \epsilon_F$). 
(a,b,c) follow from Eqs.\ (\ref{ninsnoise}) and (\ref{ninscond}) for one
channel. (d,e,f) from Eqs.\ (\ref{cond}) and (\ref{diffnoise}) are averaged 
over 
$8\times10^4$ channels. Note the pronounced resonance structure in both conductance (d) 
and noise (e), which approaches the values for the corresponding NIN 
junction at large voltages (dashed lines). The noise to current ratio (c,f) 
decays to the classical Schottky value $P/eI=1$ above the Andreev minigap.}
\label{fig2}
\end{figure}

{\it NINS junction.} In order to decribe the weak coupling of
a normal lead to an NS proximity sandwich, we consider a normal-metal -- 
insulator -- normal-metal -- superconducting (NINS) junction with a potential 
barrier of low transmission ($T\ll 1$) placed at a distance $d$ away from the 
perfect NS interface. The NINS junction serves as a model system for 
a tunneling experiment from a metallic tip to a thin film 
NS layer structure, which has permitted the observation of the Rowell-McMillan 
oscillations\cite{mcmillan}. In this elementary model, the reflection 
amplitudes $r_{\nu}\left(\epsilon\right)=
\sqrt{R}e^{i\varphi_{\nu}\left(\epsilon\right)}$ 
have a roughly constant modulus $R$ and an energy dependent phase 
$\varphi_{\nu}\left(\epsilon\right)\approx 2\left(k_{\nu}+\epsilon/v_{\nu}
\right)d + \varphi_0$, accumulated during the propagation between the potential 
barrier and the NS interface ($k_{\nu}$, $v_{\nu}$ denote the Fermi wave number 
and velocity of channel $\nu$). Consequently the shot noise (\ref{diffnoise}) 
for the channel $\nu$,
\be \frac{h}{2e^3}S_{\nu}\left(\epsilon\right) =
\frac{ 4 T^2 2R \left[1-\cos \alpha_{\nu}\left(\epsilon\right)
\right]}
{\left(T^2 + 2R \left[1-\cos \alpha_{\nu}\left(\epsilon\right)
\right] \right)^2}, \label{ninsnoise} 
\ee
depends only on the phase difference $\alpha_{\nu}(\epsilon) = 
\varphi_{\nu}\left(\epsilon\right)-\varphi_{\nu}\left(-\epsilon\right)
-2\vartheta\left(\epsilon\right)$ of electron and 
hole propagation through the X ($=$IN) region. The resonance structure of 
$S_{\nu}$ is intimately connected to the voltage (energy $\epsilon=eV$) 
dependence of the differential conductance (\ref{cond}), 
\be \frac{h}{4e^2}G_{\nu}\left(\epsilon\right) = 
\frac{ T^2}{T^2 + 2R \left[1-\cos \alpha_{\nu}\left(\epsilon\right)
\right]}. \label{ninscond} 
\ee
Figures \ref{fig2}(a) and \ref{fig2}(b) show their generic dependence on 
$\alpha_{\nu}$. 
The minima of the denominator [$\cos \alpha_{\nu}=1 \, \Rightarrow \, 
\epsilon_{\nu,n}= v_{\nu}/2d (n\pi +\arccos \epsilon_{\nu,n}/\Delta)$] 
correspond to the resonances of the Andreev quasi-bound states. 
These conductance resonances are repelled from zero voltage, since the phase 
of the Andreev reflection $\vartheta\left(\epsilon\to 0\right)=\pi/2$ 
has to be compensated by the phase difference of electron and hole propagation 
$\varphi_{\nu}\left(\epsilon\right)-\varphi_{\nu}\left(-\epsilon\right)$.
The shot noise (\ref{ninsnoise}) vanishes at these resonances 
($S_{\nu}\left(\epsilon_{\nu,n}\right)=0$). The noise maxima are found from 
$\cos\alpha_{\nu}=\left(2R-T^2\right)/2R$ at energies doubly peaked close to
the resonances. The peak separation $\delta\epsilon \sim T v_F/d$ coincides 
with the width of the resonances. 

Interestingly, this non-trivial structure
survives in the multichannel NINS junction, whose results are shown in 
Figs.\ \ref{fig2}(d) and \ref{fig2}(e). The stability of the resonance 
structure is due to its pinning to the Fermi energy. 
In contrast, the resonance structure of a potential well 
in a normal conductor is washed out in the multichannel limit. As Figures 
\ref{fig2}(b,e) show, the narrow double peak structure in the noise 
is smeared out in the multichannel limit, and the noise $S$ takes a maximum
rather than a minimum at the resonance, retracing the shape of the conductance.

At large voltages ($eV\gg v_F/d$), the shot noise approaches a constant as a 
consequence of the dephasing between the channels. The channel average may be 
evaluated by averaging over the phase $\alpha_{\nu}$ in Eqs. 
(\ref{ninsnoise}) and (\ref{ninscond}) and we obtain the 
result ($\alpha_{\nu}$ ($1/N\sum_n \to 1/2\pi \int_0^{2\pi} d\alpha$),
\be \bar{S}=\frac{1}{2\pi}\int_0^{2\pi} d\alpha\, 
S_{\nu}\left(\alpha\right) = \frac{2e^3}{h} T, \quad\quad \bar{G} = 
\frac{2e^2}{h} T. 
\ee
It follows that both the shot noise and the conductance approach the normal 
state values at voltages $v_F/d \ll eV \ll \Delta$. This limiting behavior 
demonstrates that the NIN junction is effectively decoupled from the NS 
interface in the large voltage limit and dominates both noise and conductance 
due to its low transparency. 

Let us concentrate on the overall ratio of noise power to current 
$P/eI$ 
[$P=\int_0^{e|V|} d\epsilon\, S(\epsilon)$, $I=\int_0^{eV} d\epsilon\, 
G(\epsilon)$]. This ratio has already been used successfully in determining the
unit of the charge carriers in the Fractional Quantum Hall 
Effect\cite{glattli}.
While at small voltages ($P/eI=2$) the noise carries the signature
of the Cooper pairs created by the Andreev reflection, at large voltages 
($v_F/d\ll eV \ll \Delta$) it decays to the normal state value ($P/eI=1$). 
Interestingly, this decay is immediate at the first Andreev resonance, as is 
seen from Figs.\ 2(c) and (f). In fact, in a single channel, the energy 
integration is equivalent to the integration over the phase $\alpha_{\nu}$, 
which at the midpoint between two resonances $\alpha_{\nu}$ has covered the 
period $2\pi$, thus producing $P/eI=1$ at $\alpha_{\nu}=\pi$, 
see Fig.\ 2(c). In the
transport through a normal lead weakly coupled to a NS sandwich, we thus 
find two characteristic regimes: At voltages below $eV < v_F/d$ the 
one-particle excitations in the NS structure are effectively 
gapped and we may only couple to the superconducting condensate which 
introduces the charge $2e$ in the noise power. At voltages
above the first Andreev resonance, we may couple directly to a one-particle
density of states in the normal layer and we loose the signature of the
superconductor at once.

{\sl Double barrier junction.} 
In the study of normal versus superconducting junctions, it is of interest 
to investigate the impact the quantum coherence has on the 
macroscopic limit of transport (limit of an infinite number of channels) and, 
in particular, to compare the behavior of a specific scattering region 
connected to either a normal or a superconducting reservoir.
We consider an insulator -- normal-metal -- insulator (X$=$INI) interferometer 
which we connect to normal leads in a NININ junction and to a superconducting
lead in a NINIS junction. Besides the intrinsic interest in the double barrier 
system as a paradigm, now to be combined with a superconductor, 
it may also serve as a qualitative model for a disordered conductor, due to
its strong similarity in the transmission distribution\cite{doro}. 

The resonance structure of the INI interferometer results in a 
bimodal distribution of its (overall) transmission probabilities $T$.  
For symmetric barriers it takes the form 
(infinite channel limit, $T_1=T_2\ll1$)\cite{deJong},
\be \rho\left(T\right)=\frac{1}{\pi}\frac{T_1}{2}\frac{1}{\sqrt{T^3
\left(1-T\right)}}, \quad \quad T \in \left[\frac{T_1^2}{\pi^2}, 1\right], 
\label{bimodal} \ee
which has its analog in a disordered conductor\cite{doro}. We recall that, in 
a NININ junction, the rich structure of this distribution has no impact on the 
macroscopic transport properties. The (linear) conductance and shot noise 
may be evaluated from the bimodal distribution (\ref{bimodal}), 
$\left(h/2e^2\right) G = \int dT\, \rho\left(T\right) T = T_1/2$, 
$\left(h/2e^3\right) S = \int dT\, \rho\left(T\right) T\left(1-T\right) = 
T_1/4$, and show no characteristics of coherent transport\cite{chen}. 
The conductance follows from the
series resistance of the two barriers, and an incoherent model of a double 
barrier junction gives the same suppressed noise $S/eG=1/2$ as a consequence 
of charge conservation\cite{buttbeen}.
When considering the INI interferometer in a NINIS junction instead, we find 
a non-trivial answer, both at small and large voltages. 
The {\it linear} response follows  
from Eqs. (\ref{cond}), (\ref{diffnoise}), and (\ref{bimodal}),
\ba \frac{h}{2e^2} G\left(0\right)&=&\int dT\, \rho\left(T\right) \frac{2T^2}{
\left(2-T\right)^2} = \frac{T_1}{2} \frac{1}{\sqrt{2}}, 
\nonumber \\
\frac{h}{2e^3} S\left(0\right)& = & \int dT\, \rho\left(T\right) \frac{16T^2
\left(1-T\right)}{\left(2-T\right)^4} = \frac{T_1}{2} \frac{3}{4\sqrt{2}}.
\ea
The ratio of shot noise to conductance is\cite{deJong}  
$S\left(0\right)/eG\left(0\right)=3/4$. A comparison of the shot noise to 
conductance ratios in a single barrier junction (X$=$I or X$=$IN) and a
junction with disorder (X$=$D) is instructive, see Table 1. 
In the single barrier junction the distribution of transmissions is peaked at 
$T\ll 1$. The noise is due to the Schottky type fluctuations in 
nearly closed channels, and consequently the noise ratios $S_N/eG_N=1$ in the 
NIN and $S_S/eG_S=2$ in the NIS junction differ only by the charge 
quanta involved. In a junction with disorder, the noise-to-conductance ratio is 
also doubled for the NDS case, see Table 1. However, since in the 
bimodal distribution of the transmissions the noise is produced by the channels 
with intermediate transmission $0 < T < 1$ and the current by the open channels 
with $T \to 1$, the doubling is a non-generic and thus yet unexplained 
feature. This is demonstrated by the noise-to-conductance ratio of $3/4$ in the 
NINIS as compared to $1/2$ in the NININ junction. 

At {\it finite} voltage, the shot noise is described by Eq. (\ref{diffnoise}) 
with the reflection amplitudes 
$r_{\nu}\left(\pm\epsilon\right)= r_2 + t_2^2 r_1 
e^{2ik_{\nu}\left(\pm\epsilon\right)d} /\left( 1-r_1 r_2 e^{2ik_{\nu}
\left(\pm\epsilon\right)d} \right)$ of the double barrier interferometer. 
We have evaluated this expression for a symmetric double barrier junction 
($T_1=T_2=0.05$) and display the results in 
Fig.\ \ref{fig1}. At voltages of the order of the Andreev levels 
$eV\sim v_F/d$, we find a resonance structure independent of the number of 
channels. We observe again that the differential noise follows the 
resonance peaks of the conductance. At large voltages 
($eV\gg v_F/d$), the resonances disappear as the electrons 
and holes dephase, and we expect the approach to a regime where the 
conductance and shot noise become indistinguishable from the incoherent 
addition of the NIN and NIS junctions.

In conclusion, we have expressed the differential shot noise in a disordered 
NXS junction in Eq.\ (\ref{diffnoise}) in terms of the normal and Andreev 
scattering amplitudes. We have described the resonance structure in the 
shot noise found at finite voltage in coherent transport. The 
robustness of the resonance structure in the multichannel limit is owed to 
its pinning to the Fermi energy. We have pointed out the possibility of a 
non-trivial normal versus superconducting noise ratio in the NINIS junction 
as as consequence of the bimodal distribution. 
Finally, we have found a rapid decay of double shot noise in a NINS junction 
above the Andreev minigap.

We acknowledge fruitful discussions with D. Agterberg, Ya.~M.\ Blanter, 
M.\ Dodgson, and V.~B.\ Geshkenbein.
This work has been supported by the Swiss National Science Foundation. 
G.~B.~L.\ acknowledges partial support by RFFRN 96-02-19568.
 
\begin{table}[htb]
\caption{Shot noise to conductance ratio $S/eG$ of a $NXN$ 
compared to a NXS junction; single barrier X $=$ I, double barrier X $=$ INI, 
disorder X $=$ D. The results are valid for small transmissions $T\ll 1$, 
and many channels ($N \to \infty$)\cite{khlus,deJong,chen,buttbeen}. }
\begin{tabular}{l|ccc}
      & single           & double barrier  & disorder \\ 
      & barrier $T\ll 1$ & $T_1=T_2\ll 1$ &  \\
\hline
$S_N/eG_N$ & 1     &1/2       & 1/3 \\ 
$S_S/eG_S$ & 2     &3/4       & 2/3 \\
\end{tabular}
\vspace{3mm}
\end{table}

\appendix
\section*{}

We diagonalize the BCS-Hamiltonian for superconductivity by a Bogoliubov 
transformation, considering a spatially inhomogeneous system.
The transformation is carried out by determining the quasiparticle 
wavefunctions that fulfill the Bogoliubov-de-Gennes (BdG) equations. 
We show that the symmetry in the solutions of the (BdG) 
equations with respect to reversing the energies $\epsilon_{\alpha}\to 
-\epsilon_{\alpha}$ is a consequence of the spin degeneracy. Making use of the 
spin-reversal symmetry, we express the Hamiltonian, the density, and the 
current operators in terms of the quasiparticle operators $\gamma_{\alpha}$, 
and arrive at the current expression used in Eq.\ (\ref{currentop}).

The mean-field, spin singlet BCS Hamiltonian for superconductivity 
can be expressed in the form ($h_0=\left(-i{\bf\nabla}+e{\bf A}\right)^2/2m$)
\be
{\cal H} = \int d^3x : {\bf \hat{\Psi}}^{\dag}\left({\bf x}\right) 
\left( \begin{array}{cc}
   h_0  -\mu & \Delta \\
   \Delta^* & -h_0^* + \mu \end{array} \right)
{\bf \hat{\Psi}}\left({\bf x}\right) :\, 
\label{bcs} \ee
using the Nambu spin-up annihilation operator 
\be \bf{\hat{\Psi}}\left(\bf{x}\right)= \left( \begin{array}{c}
   \hat{\Psi}_{\uparrow}\left(\bf{x}\right) \\
    \hat{\Psi}_{\downarrow}^{\dag}\left(\bf{x}\right) \end{array}  \right).
\label{Nambupsi}\ee
The pair potential is given by $\Delta({\bf x})=\lambda \langle 
\hat{\Psi}_{\downarrow}\left({\bf x}\right)\hat{\Psi}_{\uparrow}\left({\bf x}
\right) \rangle$ (coupling constant $\lambda$), the colon ($:$) denoting  
normal ordering with respect to $\hat{\Psi}_{\uparrow}$ and 
$\hat{\Psi}_{\downarrow}$. 
The Hamiltonian (\ref{bcs}) can be diagonalized by a basis transformation,
\ba 
{\bf \hat{\Psi}}\left({\bf x}\right) &=& 
\sum_{\alpha} {\bf \Phi}_{\alpha}\left({\bf x}\right) \gamma_{\alpha},   
\label{transf} \\
\gamma_{\alpha} &=& \int d^3x\, {\bf \Phi}_{\alpha}^{\dag}
\left({\bf x}\right)  {\bf \hat{\Psi}}\left({\bf x}\right),    
\label{invtransf} 
\ea 
with the eigenfunctions ${\bf\Phi}_{\alpha}\left({\bf x}\right)$ of the BdG 
equations (which follow from the insertion of the Hamiltonian 
(\ref{bcs}) into $[{\cal H},\gamma_{\alpha}]
=- \epsilon_{\alpha} \gamma_{\alpha}$),
\ba \left( \begin{array}{cc}
   h_0 -\mu & \Delta \\
   \Delta^* & -h_0^* +\mu \end{array} 
\right)
  {\bf \Phi}_{\alpha}\left({\bf x}\right) = \epsilon_{\alpha} 
{\bf \Phi}_{\alpha}\left({\bf x}\right), \label{eq:bdg} \\
{\bf\Phi}_{\alpha}\left({\bf x}\right) = \left( \begin{array}{l}u_{\alpha}\\ 
                                        v_{\alpha} \end{array} \right)  . 
\nonumber 
\ea
Note that $\gamma_{\alpha}$ are annihilation operators of spin-up states 
and thus the excitations are all described in terms of spin-up quasiparticles. 
In order to preserve the (fermionic) commutation relations we need a complete 
orthonormal set of wavefunctions of the hermitian operator in 
(\ref{eq:bdg}),
\ba \sum_{\alpha}  {\bf \Phi}_{\alpha}\left({\bf x}\right) 
{\bf \Phi}_{\alpha}^{\dag}\left({\bf x}'\right) &=& 
{\bf 1} \delta\left({\bf x}-{\bf x}'\right),  
\label{completeness} \\
\int d^3x \,{\bf \Phi}_{\alpha}^{\dag}\left({\bf x}\right) 
{\bf \Phi}_{\alpha'}\left({\bf x}\right) &=& \delta_{\alpha,\alpha'}
\label{orthonormality} 
\ea
(involving both positive and negative energy eigenstates)\cite{kuemmel}.

Consider the spin-reversal transformation $\cal S$, 
\be {\cal S} \,{\bf\hat{\Psi}} \,{\cal S}^{-1} = 
\left( \scriptstyle{\begin{array}{cc}
0 & 1 \\ -1 & 0 \end{array}} \right) 
\left({\bf\hat{\Psi}}^{\dag}\right)^{\top},
\label{spinrev} \ee
$\cal S$ linear (${\cal S}\hat{\psi}_{\uparrow}{\cal S}^{-1}\!=\!
\hat{\psi}_{\downarrow}$, ${\cal S}\hat{\psi}_{\downarrow}{\cal S}^{-1}\!=\!
-\hat{\psi}_{\uparrow}$). Noting that the order parameter $\Delta({\bf x})$ is
invariant under the transformation (\ref{spinrev}), it is easily seen that the
Hamiltonian (\ref{bcs}) is spin-reversal symmetric, $[{\cal H},{\cal S}]=0$
(this symmetry extends also to finite magnetic field, if the Zeeman splitting 
is neglected).
By means of spin-reversal we may attribute to each quasi-particle operator 
$\gamma_{\alpha}$ a linearly independent operator 
$\gamma_{\bar{\alpha}}$ through 
\be \gamma_{\bar{\alpha}}^{\dag}= {\cal S} \gamma_{\alpha} 
{\cal S}^{-1} = \int d^3x \,{\bf \hat{\Psi}}^{\dag}\left({\bf x}\right) 
\left( \scriptstyle{\begin{array}{rr} 0 & -1 \\ 1 & 0 \end{array}} \right)
{\bf \Phi}_{\alpha}^{*}\left({\bf x}\right). 
\label{relatedgamma} 
\ee
$\gamma_{\bar{\alpha}}$ describes an excitation with opposite 
energy $\epsilon_{\bar{\alpha}}=-\epsilon_{\alpha}$ 
(according to $[{\cal H},\gamma_{\bar{\alpha}}]= 
\epsilon_{\alpha} \gamma_{\bar{\alpha}}$).
From Eqs.\ (\ref{relatedgamma}) and (\ref{invtransf}) we infer the effect of 
spin-reversal on the electron-hole wavefunction,
\be 
{\bf \Phi}_{\bar{\alpha}}\left({\bf x}\right) = 
\left( \scriptstyle{\begin{array}{cc} 0 & -1 \\ 1 & 0 \end{array}} \right) 
{\bf \Phi}_{\alpha}^{*}\left({\bf x}\right) 
= \left( \begin{array}{c} -v_{\alpha}^* \\ u_{\alpha}^* \end{array} \right) .
\label{spinrevwave}
\ee
We arrive at a complete set of quasi-particle 
excitations in spin-up space, which are grouped into pairs 
${\gamma_{\alpha},\gamma_{\bar{\alpha}}}$ with energies $\pm \epsilon_{\alpha}$, as a direct consequence of spin degeneracy. Note that within the 
Nambu picture, all quasiparticles carry spin up 
[spin $\uparrow e$ and spin $\uparrow h$ (instead of spin $\downarrow e$)], 
explaining the opposite energy of the related wavefunction (\ref{spinrevwave}). 
The Hamiltonian (\ref{bcs}) takes the form,
\be {\cal H} = \sum_{\epsilon_{\alpha}>0} \epsilon_{\alpha} \left( \gamma_{\alpha}^{\dag} 
\gamma_{\alpha} + \gamma_{\bar{\alpha}} \gamma_{\bar{\alpha}}^{\dag}  -2 \int d^3x\, |v_{\alpha}|^2 \right). 
\label{Halpha}\ee
The ground state is realized by filling all the negative energy 
quasi-particle states.
 
The spin-reversal symmetry  allows us to express all equations using only half 
of the eigenstates, i.e., 
one representative ${\bf \Phi}_n$ from each pair of states 
$\left\{{\bf \Phi}_{\alpha},{\bf \Phi}_{\bar{\alpha}}\right\}$. 
In the following we choose the positive energy eigenstates, expressing the
negative energy eigenstates through (\ref{spinrevwave}). We keep 
the positive energy states for the spin-up quasiparticles ($\gamma_n=\gamma_{n\uparrow}$), 
and reinterpret the related quasi-particle states of opposite (negative) energy 
as spin-down excitations ($\gamma_{\bar{n}}=\gamma_{\bar{n}\downarrow}^{\dag}$). 
The Bogoliubov transformation (\ref{transf}) then takes the well-known form
\be {\bf\hat{\Psi}}\left({\bf x}\right) = 
\sum_n \left( \begin{array}{l}u_{n}\\  
v_{n} \end{array} \right) \gamma_{n\uparrow} 
+ \left( \begin{array}{l}-v_{n}^*\\  
u_{n}^* \end{array} \right) \gamma_{\bar{n}\downarrow}^{\dag},
\label{btstandard} 
\ee
and the Hamilton operator is expressed by
${\cal H} = \sum_{\epsilon_{n}>0} \epsilon_{n} 
( \gamma_{n\uparrow}^{\dag} 
\gamma_{n\uparrow} + \gamma_{\bar{n}\downarrow}^{\dag} 
\gamma_{\bar{n}\downarrow}  
-2 \int d^3x\, |v_{n}|^2 )$,
displaying the spin degeneracy in the usual fashion. In the same way, we give 
the density and current operators in both respresentations,
using all (indexed by $\alpha$) or only the positive energy eigenstates 
(indexed by $n$), respectively ($ u \stackrel{\leftrightarrow}
{\nabla} v = u \nabla v - (\nabla u) v$),
\end{multicols}
\ba \rho\left({\bf x}\right) &=& 
-e : {\bf \hat{\Psi}}^{\dag}\left({\bf x}\right) 
\left( \scriptstyle{\begin{array}{cc} 1 & 0 \\ 0 & -1 \end{array}} \right) 
{\bf \hat{\Psi}}\left({\bf x}\right) : 
= -e \sum_{\alpha,\alpha'} \left( 
u_{\alpha}^*({\bf x}) u_{\alpha'}({\bf x}) \gamma^{\dag}_{\alpha}
\gamma_{\alpha'} + v_{\alpha}^*({\bf x}) v_{\alpha'}({\bf x}) \gamma_{\alpha'}
\gamma^{\dag}_{\alpha} \right)
\label{rhoalpha} \\
&=& -e \sum_{\epsilon_m,\epsilon_n > 0} 
\left[u_m^*({\bf x}) u_n({\bf x}) - v_m^*({\bf x}) v_n({\bf x}) \right] 
\left( \gamma_{m\uparrow}^{\dag} \gamma_{n\uparrow} + 
\gamma_{\bar{m}\downarrow}^{\dag} \gamma_{\bar{n}\downarrow} \right) 
-2e \sum_{\epsilon_n>0} |v_n({\bf x})|^2 \nonumber \\
&& -e \sum_{\epsilon_m,\epsilon_n > 0} 
\left\{ \left[ u_m({\bf x}) v_n({\bf x}) 
+ v_m({\bf x}) u_n({\bf x}) \right] \gamma_{m\uparrow} 
\gamma_{\bar{n}\downarrow} \, + \mbox{ h.c. } \right\},
\label{rhon}
\ea
\ba 
{\bf j}\left({\bf x}\right) &=& -\frac{e}{2mi} : {\bf \hat{\Psi}}^{\dag}
\left({\bf x}\right)  {\bf \stackrel{\leftrightarrow}{\nabla} } 
{\bf \hat{\Psi}}\left({\bf x}\right) : 
= -\frac{e}{2mi} \sum_{\alpha,\alpha'} 
u_{\alpha}^*({\bf x}) {\bf \stackrel{\leftrightarrow}{\nabla} } u_{\alpha'}({\bf x}) 
\gamma^{\dag}_{\alpha} \gamma_{\alpha'} 
- v_{\alpha}^*({\bf x}) {\bf \stackrel{\leftrightarrow}{\nabla} } v_{\alpha'}({\bf x}) 
\gamma_{\alpha'} \gamma^{\dag}_{\alpha} 
\label{jalpha} \\
&=&  -\frac{e}{2mi} \left\{ \sum_{\epsilon_m,\epsilon_n > 0} 
\left(u_m^*({\bf x}) {\bf \stackrel{\leftrightarrow}{\nabla} } u_n({\bf x}) 
+ v_m^*({\bf x}) {\bf \stackrel{\leftrightarrow}{\nabla} } v_n({\bf x}) 
\right) \left( \gamma_{m\uparrow}^{\dag} \gamma_{n\uparrow} 
+ \gamma_{\bar{m}\downarrow}^{\dag} \gamma_{\bar{n}\downarrow} \right) 
-2 \sum_{\epsilon_n>0} v_n^*({\bf x}){\bf \stackrel{\leftrightarrow}{\nabla} }
 v_n({\bf x}) \right\} \nonumber \\
&& -\frac{e}{2mi} \sum_{\epsilon_m,\epsilon_n > 0} \left\{ 
\left(- u_m({\bf x}) {\bf \stackrel{\leftrightarrow}{\nabla} } v_n({\bf x}) 
+ v_m({\bf x}) {\bf \stackrel{\leftrightarrow}{\nabla} } u_n({\bf x}) \right) 
\gamma_{m\uparrow} \gamma_{\bar{n}\downarrow} \, - \mbox{ h.c. } \right\} .
 \label{jn} 
\ea
\widetext
\begin{multicols}{2}
The current operator is easily generalized to finite magnetic field. We note
that current and density operators obey the continuity equation
\be \frac{\partial\rho({\bf x})}{\partial t} + {\bf \nabla \cdot j}({\bf x}) = 
2ie {\bf \hat{\Psi}}^{\dag}\left({\bf x}\right) 
\left( \scriptstyle{\begin{array}{cc} 0 & \Delta({\bf x}) \\ -\Delta^*({\bf x}) & 0 
\end{array}} \right)  {\bf \hat{\Psi}}\left({\bf x}\right),
\ee
where the right hand side vanishes when taking the (self-consistent) 
expectation value. The effective current expression used as a starting point
in Eq. (\ref{currentop}) is obtained from (\ref{jn}), accounting for the spin 
degeneracy by a factor 2 and dropping the particle non-conserving last term, 
which is eliminated by the frequency integration in 
(\ref{power}).

\end{multicols}


\begin{thebibliography}{99}
\bibitem{notation}We use the notation of Ref.\ \cite{leso}, with the only 
change, that the global reflection amplitudes are denoted by lower-case instead of 
capital letters, e.g., $r_{he}$ instead of $R_{he}$.
\end{thebibliography}
\end{document}